\documentclass[12pt,preprint]{aastex}
\usepackage{natbib}

\def\hst{{\it HST\/}}
\def\ro{{\it ROSAT\/}}
\def\chandra{{\it Chandra\/}}
\def\rxja{RX~J1242.6--1119A}
\def\rxjb{RX~J1624.9+7554}
\def\ngc{NGC~5905}
\slugcomment{{\it Received 2003 October 31; accepted 2003 December 9}}

\shorttitle{Tidal Disruption Events}
\shortauthors{Halpern et al.}

\begin{document}

\title{Follow-Up \chandra\ Observations of Three Candidate Tidal Disruption Events}

\author{J. P. Halpern, S. Gezari}
\affil{Department of Astronomy, Columbia University, 550 West 120th Street, New York, NY 10027-6601}
\email{jules@astro.columbia.edu}

\author{S. Komossa}
\affil{Max-Planck-Institut f\"ur extraterrestrische Physik, Giessenbachstrasse 1, D-85748 Garching, Germany}

\begin{abstract}
Large-amplitude, high-luminosity soft X-ray flares were detected by the \ro\
All-Sky Survey in several galaxies with no evidence of Seyfert
activity in their ground-based optical spectra.  These flares 
had the properties predicted for a tidal disruption of a star
by a central supermassive black hole.
We report \chandra\ observations of three of these galaxies
taken a decade after their flares that reveal weak nuclear X-ray sources
that are from 240 to 6000 times fainter
than their luminosities at peak, supporting the theory that these were 
special events and not ongoing active galactic nucleus (AGN) variability.
The decline of \rxjb\ by a factor of 6000
is consistent with the $(t-t_D)^{-5/3}$ decay predicted for the 
fall-back phase of a tidal disruption event, but only if \ro\ was lucky
enough to catch the event exactly at its peak in 1990 October.
\rxja\ has declined by a factor of 240, also consistent with $(t-t_D)^{-5/3}$.
In the \ion{H}{2} galaxy \ngc\ we find only resolved, soft X-ray
emission that is undoubtedly associated with starburst activity. 
When accounting for the starburst component, the \ro\
observations of \ngc, as well as the \chandra\ upper limit on its nuclear
flux, are consistent with a $(t-t_D)^{-5/3}$ decay by at least a factor of 1000.
Although we found weak Seyfert~2 emission lines in {\it Hubble Space Telescope}
spectra of \ngc, indicating that a low-luminosity AGN was present
prior to the X-ray flare, we favor a tidal disruption explanation for the
flare itself.
 
\end{abstract}


\section{Introduction} \label{intsec}

Dormant, supermassive black holes, suspected to be present in the
centers of many normal galaxies, should reveal themselves by a UV/X-ray
flare when they tidally disrupt a star and some fraction of the stellar
debris is accreted.  Tidal disruption flares were proposed by \cite{li79}
and Rees (1988, 1990) as a probe for supermassive black holes
in the centers of inactive galaxies.  As argued by \cite{ul99}, the spectrum of
such a flare will be characterized by the 
blackbody temperature of a thick disk or spherical envelope at the tidal radius,
$T_{\rm eff} \approx (L_{\rm Edd}/4\pi\sigma R_T^2)^{1/4} = 3.7 \times 10^5 M_8^{1/12}$~K
for a solar-type star.
The flare would begin when the most tightly bound portion of the tidal debris
returns to the pericenter of the star's orbit and accretes onto the black hole.
This first return would occur at a time $t_0$ following the disruption at $t_D$,
where $(t_0 - t_D) \sim 1.1 M_8^{1/2}$ yr.
This estimate applies to a non-rotating star; in the likely case
that the star is spun up to near break-up before disruption, $t_0 - t_D$
is reduced by the factor $3^{-3/2}$ (e.g., Li, Narayan, \& Menou 2002).
The maximum return rate of debris according to numerical simulations (Evans \& Kochanek 1989) is
$\dot M_{\rm max} \sim 0.14\,M_{8}^{1/2}\,M_{\sun}$ yr$^{-1}$
and occurs at a time $(t_{\rm max} - t_D) \sim 1.5\,(t_0 - t_D)$.
After the peak of the flare, material returns at the declining rate
$\dot M(t) = 0.3\,M_8\,[(t-t_D)/(t_0-t_D)]^{-5/3}\,M_{\sun}$
yr$^{-1}$, which is an important factor that controls the decay of the
luminosity over the next few
years. For $M_{\rm BH} < 10^{7}\ M_{\sun}$, the maximum return rate is super-Eddington,
resulting in a flare with
$L_{\rm flare} \geq \eta \dot M_{\rm Edd} c^{2} > 1.3 \times 10^{45} M_7$ ergs s$^{-1}$.  
Thus, for low-mass central black holes ($M_{\rm BH} < 10^7 M_{\sun}$), tidal disruption
theory predicts luminous flares of up to $10^{45}$ ergs s$^{-1}$, 
peaking in the soft X-ray domain, with time scales on the order of months.
For $M_{\rm BH} > 2 \times 10^{7}\ M_{\sun}$, stellar debris takes a longer
time to fall back than the time scale on which it can circularize and radiate;
in this case accretion probably proceeds through a thin disk \citep{ul99}.

The \ro\ All-Sky Survey (RASS; Voges et al. 1999) conducted
in 1990--1991 was an ideal experiment to detect these
flares since it sampled hundreds of thousands of galaxies in the soft X-ray band.
\ro\ detected soft X-ray outbursts from several galaxies with no previous evidence
of Seyfert activity (see Komossa 2002 for a review).   From the statistics
of the RASS, \cite{do02} calculated a rate of $\approx 1 \times 10^{-5}$ yr$^{-1}$
for X-ray flares from these non-active galactic nucleus
(non-AGN) galaxies, consistent with expected rates
\citep{ma99,sy99,wm03}.
In order to test in an independent way whether
these flares were in fact tidal disruption events, as opposed to some other form of 
extreme AGN variability, \cite{ge03} obtained optical spectra
of three of these non-AGN galaxies with the {\it Hubble Space Telescope}
(\hst ) through narrow slits to search for persistent Seyfert activity.
These spectra were up to a factor of 100 more sensitive to nuclear 
activity than previously obtained ground-based data.
Two of the galaxies, \rxja\ \citep{ko99} and \rxjb\
(Grupe, Thomas, \& Leighly 1999),
showed no evidence of emission lines or a non-stellar continuum in
their \hst\ nuclear spectra, consistent with their ground-based
classification as inactive.  On the other hand, \ngc\ \citep{ba96},
a starburst galaxy with strong emission lines,
was found by \cite{ge03} to have in its inner $0.\!^{\prime\prime}1$
a nucleus with narrow emission line ratios
indicating a Seyfert~2 classification.
This weak Seyfert~2 nucleus requires a low level of prior non-stellar
photoionization powered by accretion, which raises some doubt about
whether its X-ray flare must have been a tidal disruption event
but does not rule it out.
In this paper, we report on follow-up X-ray observations with \chandra\
of \ngc, \rxja, and \rxjb.  The superb spatial resolution of
\chandra\ enables an even more sensitive search for nuclear X-ray activity
in these now very weak X-ray sources, and resolves any non-nuclear sources of
X-ray emission, both of which are needed to test more rigorously the tidal 
disruption hypothesis.

\section{Observations and Basic Results} \label{obssec}

All three targets were observed with the Advanced CCD Imaging Spectrometer
(ACIS; Burke et al. 1997) on board the \chandra\ X-Ray Observatory
\citep{we96}.  In each case the galaxy was
positioned on the back-illuminated S3 chip of the ACIS-S array.
We used the standard processed and filtered
event data with the latest aspect alignments, with the
exception that the 0.5 pixel ($0.\!^{\prime\prime}25$) randomization that is ordinarily
applied to the photon positions was reversed, restoring slightly sharper
images.  The $0.\!^{\prime\prime}5$ ACIS pixels slightly undersample the
on-axis point-spread function of the \chandra\ mirrors
in the restored images.
Table~\ref{tbl1} is a summary log of the \chandra\ observations and basic results.
Since all of the X-ray sources are too weak to apply spectral fits, we use
their count rates and reasonable assumptions about an appropriate spectral
model to estimate their luminosities using the Web-based simulator
PIMMS.\footnote{Available at http://asc.harvard.edu/toolkit/pimms.jsp.}
In order to account approximately for time-dependent degradation of the
ACIS throughput below 1~keV, we used the PIMMS setup for the AO4 observing
period (2002--2003).   Systematic errors in luminosity associated with this
choice will be of order $10\%$.  We quote absorbed
fluxes and unabsorbed luminosities in the 0.2--2.4~keV band using
$H_0 = 75$ km~s$^{-1}$~Mpc$^{-1}$.  Since in many cases published
\ro\ luminosities are quoted for the 0.1--2.4~keV band and
$H_0 = 50$ km~s$^{-1}$~Mpc$^{-1}$, we make the required conversions
where necessary.

\begin{figure}
\vskip -0.9in
\plotone{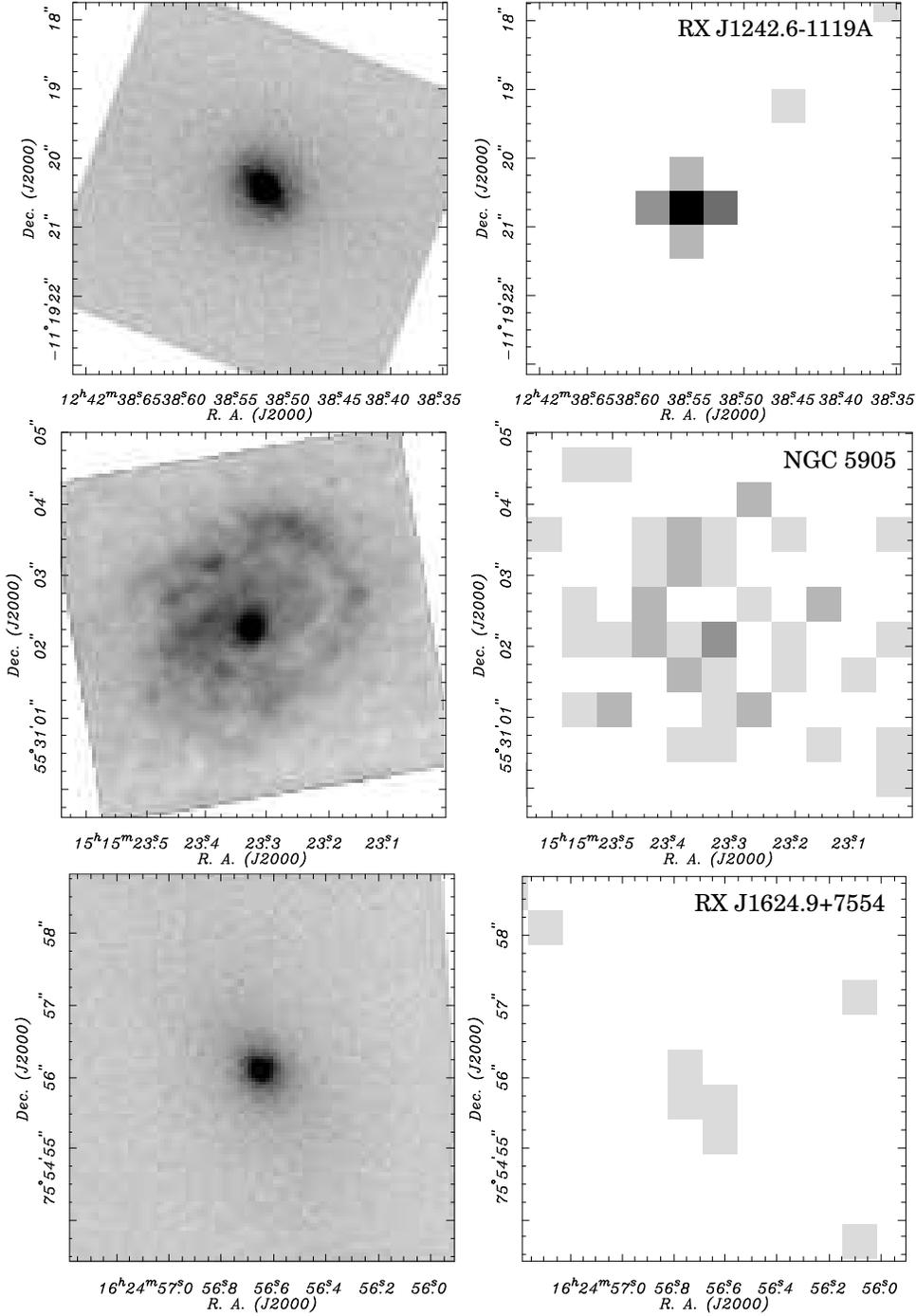}
\vskip -0.25in
\caption{\hst\ acquisition images
($5^{\prime\prime} \times 5^{\prime\prime}$)
of the target galaxy nuclei ({\it left}),
compared with \chandra\ ACIS-S3
images at the same scale ({\it right}).
The gray scale in the \chandra\ images runs from zero 
photons ({\it white}) to seven photons ({\it black})
per $0.\!^{\prime\prime}5 \times 0.\!^{\prime\prime}5$
pixel. The galaxy \rxja\ clearly hosts a point-like 
nuclear X-ray source (18 photons), 
while the X-ray emission from NGC~5905 (48 net photons) is mostly if not 
entirely diffuse, and distributed similarly to the inner spiral structure.
The nature of the four photons coincident with the nucleus of \rxjb\ is not 
clear.\label{image1}}
\end{figure}

For illustrative comparison, we also use images of the nuclei of these galaxies
that were obtained by the Space Telescope Imaging Spectrograph
(STIS) CCD in the course of target acquisition for our
spectroscopy program reported in \cite{ge03}.
These consist of small, $5^{\prime\prime} \times 5^{\prime\prime}$
windows that were exposed through the long-pass filter F28$\times$50LP,
yielding a broad band-pass from $5500$ \AA\ to 1 $\mu$m.  Figure~\ref{image1} shows the STIS images side by side with the \chandra\ image of each target.

Since the default astrometric calibration of \hst\ images is generally
not as accurate as that of \chandra, we used ground-based images to
derive optical positions of the nuclei of the targets with respect to the
astrometric grid of USNO-A2.0 stars \citep{mo98}.  Optical positions so
derived are expected to be accurate to $\approx 0.\!^{\prime\prime}3$, the
typical uncertainty of the USNO-A2.0 astrometry.  We then registered
the STIS images to these astrometric coordinates using the
nuclei of the target galaxies.  Since \chandra\ aspect
reconstruction is known to have a random error of only $\approx 0.\!^{\prime\prime}6$
at $90\%$ confidence, we expect that X-ray and optical positions so derived 
will agree to $\approx 1^{\prime\prime}$ or better even before applying
corrections that can be made by optically identifying serendipitous X-ray sources.
It was possible to meaningfully check the \chandra\ aspect solution using
multiple serendipitous sources in only one case.
The results are described below.

\subsection{NGC 5905}

\begin{figure}
\plotone{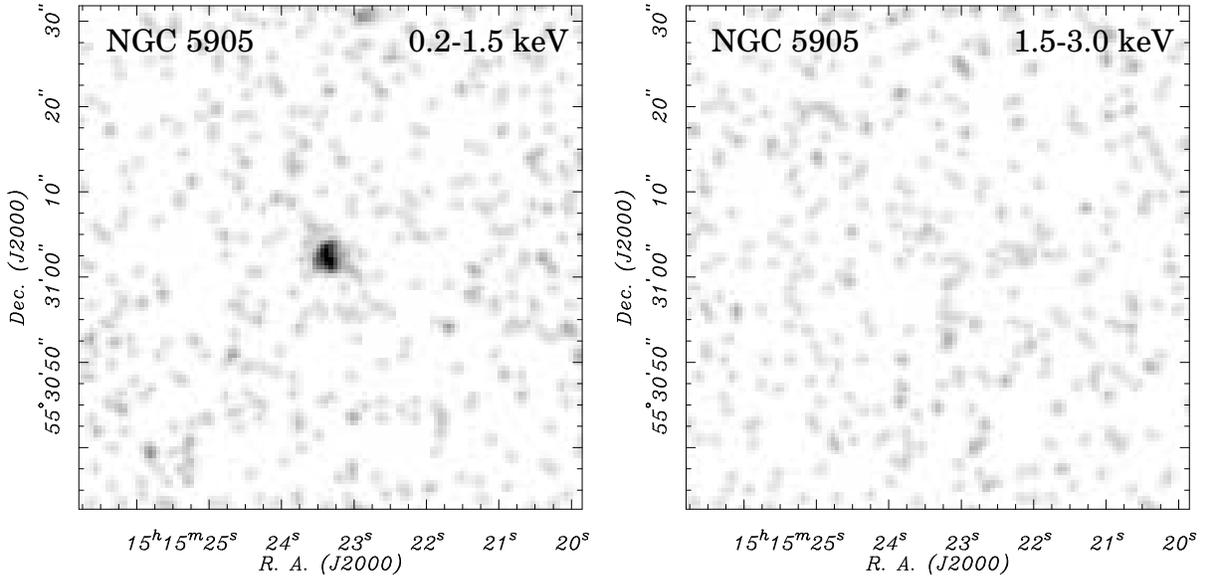}
\vskip -4.0in
\caption{\chandra\ ACIS image of the
central region of NGC 5905, smoothed with a Gaussian
of $\sigma = 1^{\prime\prime}$.  The absence of emission
above 1.5~keV argues against an AGN and is interpreted as
a diffuse starburst source.\label{image2}}
\end{figure}

The \chandra\ image of \ngc\ consists of a diffuse source with a diameter
of $\approx 4^{\prime\prime}$ that is coincident with the nucleus of the
galaxy.  Figure~\ref{image1} shows that the distribution of X-ray photons
is consistent with the pattern of the inner spiral structure seen in the
\hst\ image.  After background subtraction, $\approx 48$ photons are
detected in this region.  Furthermore, Figure~\ref{image2} indicates that
all of this emission is confined to energies below 1.5~keV.
\hst\ spectra in this region \citep{ge03} have strong Balmer
emission lines, indicating that the spiral structure is dominated by
young stars and \ion{H}{2} regions.  Since it is likely that the X-rays
originate from processes specific to starbursts, such as O star
winds, superbubbles, and perhaps old supernova remnants, we use a Raymond-Smith
thermal plasma model to estimate the flux and luminosity of this source.
As indicated in Table~\ref{tbl1}, we find that for temperatures around
$5\times 10^6$~K, the X-ray luminosity is $4.4 \times 10^{39}$ ergs~s$^{-1}$
in the 0.2--2.4 keV band, or $4.9 \times 10^{39}$ ergs~s$^{-1}$
in the 0.1--2.4 keV band.
The absence of hard X-rays argues against a large contribution of
X-ray binaries to this flux, although a few can be present.
Diffuse sources of similar luminosity and
extent have been detected in the nucleus and bar of NGC~1672
and in other nearby starburst galaxies \citep{br96,de00}.

Although the brightest pixel in the X-ray image falls on the center of the
galaxy, it contain only three photons, and does not constitute strong evidence
of an active nucleus, whether from ongoing Seyfert
activity or from the tail of the flare.   Both the softness of the source and
the lack of a significant central peak lead us to believe that most of
it is starburst emission.  Nevertheless, we derive a conservative upper
limit to the flux of a nuclear X-ray source by assigning the central pixel
and its immediate neighbors, a total of eight photons,
to the upper limit.  If modeled as a power law of photon index
$\Gamma = 2.5$, this corresponds to less than $9.1 \times 10^{38}$ ergs~s$^{-1}$
in the 0.2--2.4 keV band, or less than $1.4 \times 10^{39}$ ergs~s$^{-1}$
from 0.1--2.4 keV.
If instead we assume a blackbody of $kT = 0.06$~keV, similar to the
spectrum of the \ro\ flare, then the upper limit is less than
$2.6 \times 10^{39}$ ergs~s$^{-1}$ in the 0.1--2.4 keV band.

\subsection{RX J1624.9+7554}

\begin{figure}
\vskip -2.0 truein
\plotone{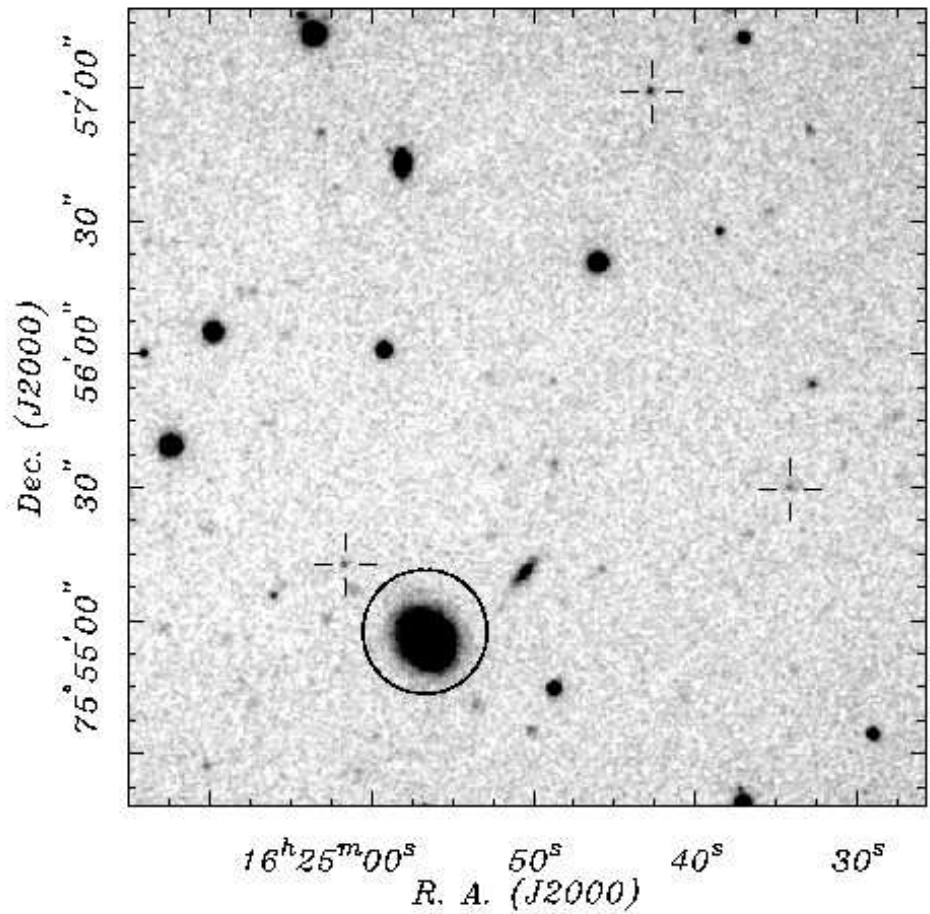}
\caption{$R$-band image of the region around \rxjb\ from
the MDM 2.4~m telescope.  The circle is the original
\ro\ $2 \sigma$ error location of \rxjb\ from \cite{gr99},
and the crosses are the locations of \chandra\ 
serendipitous sources,
the properties of which are listed in Table~2.\label{image3}}
\end{figure}

The original \ro\ error circle of this source contains a single galaxy that
was studied optically by \cite{gr99} and \cite{ge03}.
Of the three targets studied here, \rxjb\ has the weakest X-ray detection.
It is not clear whether
the four photons detected by \chandra\ from the vicinity
of its nucleus in Figure~\ref{image1} are related to the original flare,
or even whether they are coming from the optical nucleus.
However, since their coordinates coincide to
within $0.\!^{\prime\prime}65$, we assume that the X-ray source represents
a weak detection of the nucleus of \rxjb.
Three additional point sources that are seen
by \chandra\ on the ACIS-S3 CCD can be used to verify the X-ray astrometry.
Figure~\ref{image3}, an $R$-band CCD image that was obtained
on the MDM 2.4~m telescope on 1999 March 8, shows the locations of the
detected sources, all of which coincide with faint objects in the magnitude
range 21.2--21.6, and are possibly QSOs or other types of AGNs.  Their X-ray
and optical positions, listed in Table~\ref{tbl2}, agree on average to within
$0.\!^{\prime\prime}3$ in each coordinate.  Thus, we have verified the X-ray
astrometry and we do not make any further adjustments to it.
The closest X-ray source to \rxjb\ is $25^{\prime\prime}$ away,
nominally too far from the \ro\ error circle ($2 \sigma$ radius of
$14^{\prime\prime}$) to be identified with the \ro\ flare.
If by unfortunate coincidence this 12 photon source CXOU J162501.6+755512
is actually the source of the \ro\ flare, then we should consider the
possibility that it is a variable Galactic object such as an AM~Her star.
However, we have carefully examined the RASS photons from \rxjb, confirming
the previously published source position.  We accept the original
optical identification of the host galaxy of the X-ray flare as most likely,
and proceed to consider the implications of the weak \chandra\
``detection'' of the same galaxy.

The measured energies of the four ``nuclear'' photons range from
0.7 to 4.8 keV.  This range would not be expected for a soft blackbody
or diffuse thermal source.  Therefore,  
we treat this as a single point source and estimate its
luminosity assuming a power-law spectrum of photon index
$\Gamma = 2.5 \pm 0.2$.  The result is $(1.7 \pm 0.8) \times 10^{40}$
ergs~s$^{-1}$ in the 0.2--2.4~keV band,
or $(2.7 \pm 1.3) \times 10^{40}$ ergs~s$^{-1}$ from 0.1--2.4~keV.
Given the uncertainty about the existence or location
of this source, it is perhaps safer to regard it as an upper limit to
the persistent X-ray luminosity of \rxjb.  This observation extends the
total amplitude of X-ray variability of \rxjb\ to a factor of $\approx 6000$,
since its originally detected luminosity was $\approx 1.6 \times 10^{44}$
ergs~s$^{-1}$ \citep{gr99}.

\subsection{RX J1242.6--1119A}

\rxja\ is the only one of the three \chandra\ targets that has a clear
point X-ray source associated with its nucleus.  The X-ray and optical positions
differ by $0.\!^{\prime\prime}5$, which is not significant.
The centering of the X-ray source on the optical nucleus of \rxja\
also tends to rule out the fainter companion galaxy 
RX~J1242.6--1119B as the source of flare; their positions were
not distinguished by \ro\ \citep{ko99}.
The measured energies of the 18 X-ray photons range from
0.4 to 4.6~keV.  This range would not be expected for a soft blackbody
of $kT = 0.06$ keV, similar to the original flare spectrum.
Moreover, an observation of \rxja\ by {\it XMM}-Newton in 2001 June
\citep{ko04} can be fitted by a power law of photon index
$\Gamma = 2.5 \pm 0.2$.  Therefore, we estimate the source
luminosity assuming such a power law.  The
result is $(1.1 \pm 0.3) \times 10^{41}$ ergs~s$^{-1}$
in the 0.2--2.4~keV band, or $(1.7 \pm 0.5) \times 10^{41}$ ergs~s$^{-1}$
from 0.1--2.4~keV.  This detection extends the
total amplitude of X-ray variability of \rxja\ to a factor of $\approx 240$,
since its originally detected luminosity was $\approx 4 \times 10^{43}$
ergs~s$^{-1}$ \citep{ko02}.

\section{Interpretation} \label{discsec}

\subsection{NGC 5905}

\ngc, the galaxy with the best-sampled historical X-ray light curve,
shows a fading of the flare luminosity at a rate close to the predicted
accretion rate $\dot M(t) \propto t^{-5/3}$ (Komossa \& Greiner 1999),
which \cite{li02} regard as strong evidence 
that its flare was a tidal disruption event.
The \chandra-detected diffuse X-ray emission around the nucleus of
\ngc\ is comparable in luminosity to the lowest state measured in
the final \ro\ observation in late 1996.  Correcting the luminosity
listed in Table~\ref{tbl1} to the 0.1--2.4 keV band and to the
$H_0 = 50$ km~s$^{-1}$~Mpc$^{-1}$
used by \cite{kb99}, we find that the \chandra-measured
flux can account for $0.66 \pm 0.17$ of the minimum
flux measured by \ro.  Considering the different band passes and 
spatial resolution of the two instruments, as well as the lack of
constraints on the appropriate spectral model, we consider this
fraction to be consistent with unity.

\begin{figure}
\plotone{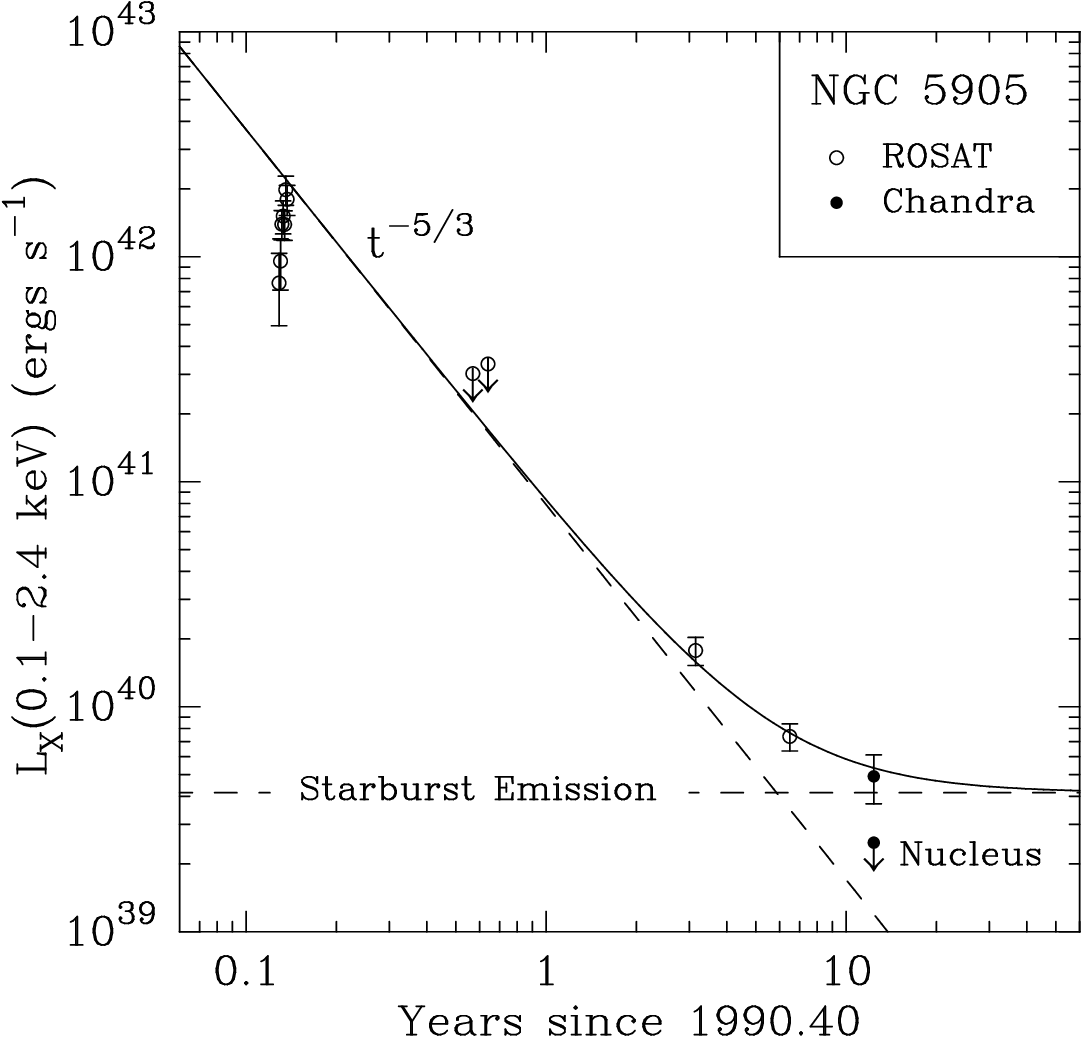}
\caption{X-ray measurements of \ngc\
from \ro\ \citep{ba96,kb99} and \chandra\ (this work).
Luminosities assume a $kT = 0.06$~keV
blackbody spectrum for the nuclear source,
and have been adjusted to a common unabsorbed (0.1--2.4~keV)
energy band and $H_0 = 75$ km~s$^{-1}$~Mpc$^{-1}$.  The resolution
of the starburst X-ray component by \chandra\ allows a good fit 
({\it solid line)} to a $(t-t_D)^{-5/3}$ decay plus a constant
({\it dashed lines}). The fitted epoch of the tidal disruption event is
$t_D=1990.40$, $\approx 50$ days before the peak of the flare caught
by \ro.\label{graph1}}
\end{figure}

Figure~\ref{graph1} shows a history of the X-ray luminosity of \ngc\
from \ro\ \citep{ba96,kb99} and \chandra.  Luminosities in this Figure
assume a $kT = 0.06$~keV blackbody spectrum for the nuclear source,
and have been adjusted to a common (0.1--2.4~keV) energy band
and $H_0 = 75$ km~s$^{-1}$~Mpc$^{-1}$.  Note that the resolution
of the starburst X-ray component by \chandra\ allows a good fit 
to a $(t-t_D)^{-5/3}$ decay plus a constant.  Taking the
starburst component into account improves the fit of the
measurements and upper limits to the $(t-t_D)^{-5/3}$ line over
the fits made by \cite{kb99} and \cite{li02}.  The \chandra\
upper limit on the nuclear flux is also consistent with such a decay,
and by virtue of its sensitivity extends the total observered
amplitude of the flare to a factor of $10^3$, which is a larger
range than has been seen in any AGN.

It is important to note that these data do not tightly constrain the
{\it value} of the decay index, since the actual time of the tidal disruption
event $t_D$ is unknown.  Rather, a decay index of $-5/3$ is assumed and
fitted to the highest flux point observed during the flare and all
of the subsequent detections.  An acceptable fit is found provided that $t_D=1990.40$,
which is $\approx 50$ days before the peak of the luminosity seen by \ro.  
If the RASS observation did indeed catch the peak of the flare,
it is quite constraining of $t_D$, perhaps fortuitously so since
the observation was only 4 days long.  Our estimate of $t_D$ differs
by only about 10 days from that of \cite{li02}, which is within their
assumed margin of error.
\cite{li02} showed that in the case of a star that
is spun up to near break-up,
this delay is consistent with the expected time for bound material
to return to a black hole of $< 10^8\,M_{\odot}$, which is consistent
with the upper limit of $< 1.7 \times 10^8\,M_{\odot}$ derived
by \cite{ge03} from the H$\alpha$ emission-line velocities
in the \hst\ spectrum.
 
\cite{li02} also argued that
since the flare luminosity is much less than $L_{\rm Edd}$, it is likely
that only a small fraction of a solar mass was accreted, perhaps from a
brown dwarf or just the outer layers of a low-mass star.
The integral of the luminosity in the $t^{-5/3}$ decay from the
presumed peak at $t_{\rm max} = 1990.53$ to $t = \infty$
is $\approx 1.3 \times 10^{49}$ ergs,
with a smaller amount coming before the peak.  This
requires an accretion of only $\approx 8 \times 10^{-6}\eta^{-1}\,M_{\odot}$,
where $\eta$ is the efficiency of converting mass to energy.

Even in the absence of other evidence for Seyfert activity,
the narrow emission-line ratios found in the nucleus of \ngc\
by \hst\ require excitation by a non-stellar ionizing continuum.
\cite{ge03} argued that these lines are unlikely to have been
excited by the X-ray flare, but indicate prior, ongoing activity.
In an erratum to \cite{ge03}, a revised nuclear H$\alpha$ luminosity
of $1.3 \times 10^{38}$ ergs s$^{-1}$ was measured.  This quantity
can be used to predict the time-averaged soft X-ray luminosity of the 
nucleus using the correlation of 0.1--2.4~keV
luminosity with H$\alpha$ luminosity derived by
\cite{ha01} from observations of low-luminosity AGNs.
They find $L_{\rm X} = 7\,L_{{\rm H}\alpha}$ on average.
The predicted soft X-ray luminosity of NGC 5905 is then
$\sim 9 \times 10^{38}$ ergs s$^{-1}$, which is comparable to the
\chandra\ upper limit of $< 1.4 \times 10^{39}$
ergs s$^{-1}$ on the nuclear luminosity in the 0.1--2.4~keV band.
Considering the scatter in this relation, we are unable to rule out the
existence of a normal, low-luminosity AGN in the nucleus of \ngc\ that
survived after the X-ray flare.  Given the huge amplitude of the flare,
we favor a tidal disruption explanation for its origin
even if there is also an underlying low-luminosity AGN.
Deeper exposures with \chandra\ will
be able to test for a persistent AGN, i.e., by detecting a positive excess
above the extrapolated $t^{-5/3}$ decline shown in Figure~\ref{graph1}.
The present upper limit from a short observation is not even a factor of
2 above the extrapolation.
 
\subsection{\rxjb\ and \rxja }

\rxjb\ was detected only once by \ro, and we cannot fit a power-law
decay to it because there is no information about the time of origin $t_D$
of the flare.  If we {\it assume} that the factor of $6000$
observered X-ray decline follows a $t^{-5/3}$ law, then the time
of peak emission $t_{\rm max}$ would have occurred only $\approx 24$
days before the
RASS observation in 1990, which is fortuitous timing similar to the
\ngc\ discovery.  (Of course, it is the all-sky coverage of the RASS 
that enabled it to detect these rare events).  However,
the RASS observation itself lasted 8.5 days \citep{gr99},
during which time the light curve showed only random 
variability around a mean count rate.  This constancy can be
understood in the simplest decay scenario only if the RASS observation
was timed exactly to catch the flare at its peak, when it had
stopped rising but had not yet entered a $t^{-5/3}$ decline phase.
Alternatively, the constant \ro\ count rate may be interpreted
as an Eddington-limited phase in the flare, in which case
$L_{\rm X} \approx 1.6 \times 10^{44}$ implies $M_{\rm BH} \sim 10^6\,M_{\sun}$.
The duration of such a phase is $\sim 0.76 M_7^{-2/5}$~yr \citep{ul99},
or $\sim 2$~yr; under this interpretation the
timing of the RASS observation was not special.
The total energy in the RASS light curve alone is
$\approx 1 \times 10^{50}$ ergs.  While this is less than that
observed in many gamma-ray burst (GRB) afterglows, the long constant
phase and soft X-ray spectrum are dissimilar to all well-observered
GRBs, and do not support such an identification.

The only other indication of whether the observed
factor of 6000 X-ray decline in \rxjb\ can be a property of a persistent
AGN comes from the absence of optical activity in its
\hst\ spectrum \citep{ge03}.  An upper limit to the flux of a narrow
H$\alpha$ emission line of $ < 4 \times 10^{-17}$ ergs cm$^{-2}$ s$^{-1}$
can be derived from that spectrum, corresponding to a luminosity of
$ < 3.1 \times 10^{38}$ ergs s$^{-1}$.  Narrow emission lines are
more useful for this test than broad emission lines because they
are powered by the average ionizing flux over decades, rather than over
days for broad emission lines.  The H$\alpha$ upper limit is a factor of 
90 less than the present $(2.7 \pm 1.3) \times 10^{40}$ ergs~s$^{-1}$
X-ray luminosity, compared to the \cite{ha01}
average ratio $L_{\rm X}/L_{{\rm H}\alpha}=7$,
which would tend to argue that there is {\it not\/}
an ordinary, persistent AGN with such an X-ray luminosity in \rxjb.  Additional,
deeper X-ray observations during the coming years can clarify the nature
of the low state of this galaxy nucleus.

Similarly to the case  of \rxjb, \rxja\ was observed in a high state only
once.   However, the high state of \rxja\ was observed in pointed mode
1.6~yr {\it after} a non-detection in the RASS during
1990 December -- 1991 January \citep{ko99}.
Therefore, we are able to delimit the time of the presumed tidal disruption
event to later than 1991.0 and earlier than 1992.7.  If the one \ro\ detection
at $\approx 4 \times 10^{43}$ ergs~s$^{-1}$ and the one \chandra\ detection 
at $\approx 1.7 \times 10^{41}$ ergs~s$^{-1}$ are forced to fit a $t^{-5/3}$ law,
then $t_D \approx 1991.36$, which is consistent with the non-detection in
the RASS.  A caveat is needed here.  These luminosities were calculated 
assuming different spectra models (soft blackbody for the RASS {\it vs.}
power-law for \chandra) because the spectrum appears harder in the latter case.
It is not necessary that the soft X-ray band contains the bulk of
the bolometric luminosity at late times if the emission is no longer optically thick
at low accretion rates.
The theoretical $t^{-5/3}$ decline applies to the {\it accretion rate}, whereas
we may expect a departure of the {\it X-ray flux} from this power law
as the emission spectrum broadens.  If such an effect is operating, the X-rays
underestimate the bolometric luminosity, and the inferred disruption time $t_D$
should be revised to an earlier date.
As in the case of \rxjb, it is not excluded that the flare from
\rxja\ was observed in an Eddington-limited phase and that $M_{\rm BH}$ is
therefore relatively small.

As we did for the other objects, we can use the \hst\ spectrum of \rxja\ to quantify
and interpret the absence of AGN emission lines.
Similarly to \rxjb, we find that the upper limit to the flux of a narrow
H$\alpha$ emission line is $ < 4 \times 10^{-17}$ ergs cm$^{-2}$ s$^{-1}$,
corresponding to $L({\rm H}\alpha) < 2 \times 10^{38}$ ergs s$^{-1}$.
Since this limit is a factor of 850 less than the present X-ray luminosity,
it appears that a ``normal'' AGN with this X-ray luminosity is not
a persistent feature of \rxja\ either.  Future observations with \chandra\
will be able to track the expected further decay of the X-ray source
in \rxja\ according to the tidal disruption scenario.

\section{Conclusions} \label{dissec}

\chandra\ observations of \ngc, \rxja, and \rxjb\ in 2001 and 2002
show that their X-ray
fluxes are continuing to decline at a rate that is consistent with the
predicted accretion rate as a function of time in the fall-back phase
of a tidal disruption event. \ngc\ and \rxjb\ are observered to be
factors of 1000 and 6000 fainter, respectively, than their peak
luminosities detected in the RASS.   Only \rxja\ still has an identifiable
nuclear X-ray source, whose luminosity is nevertheless a factor of 240
less than its peak in 1992.  Since \rxja\ and \rxjb\ were confirmed
to be inactive galaxies from the absence of broad or narrow emission lines,
or non-stellar continuum in their \hst\ spectra, the most natural
interpretation of their presently weak X-ray emission is the continuing
decline following the tidal disruption of a star by an otherwise dormant
central supermassive black hole.

The relatively low peak luminosity and fluence of the X-ray flare in
\ngc\ is most simply interpreted as the accretion of only
$\sim 10^{-4}\,M_{\sun}$ of stellar debris, while the short duration of the
peak is consistent with $M_{\rm BH} \leq 1 \times 10^8\,M_{\sun}$.  The much
higher luminosities of the flares in \rxja\ and \rxjb\ do not necessarily
imply larger black hole masses for those objects, since a super-Eddington 
infall rate onto a smaller black hole can last for of order a year
and maintain a constant luminosity for that time.  A longer peak duration
would make it easier to understand why the RASS caught all of these events
at their maximum flux.  \cite{wm03} argued from theory that the rate of
tidal disruption in galaxy nuclei should actually increase with decreasing
$M_{\rm BH}$; if so, this should be considered seriously as a selection effect.

In the case of \ngc, the \hst\ detection of weak, Seyfert~2 emission lines
in its nucleus raises additional uncertainty about
whether its X-ray flare was the result of a tidal disruption or just
exceptionally high-amplitude variability of its low-luminosity Seyfert nucleus
that is not yet explained.
The ability of \chandra\ to resolve the starburst source in \ngc\ enabled
a further order-of-magnitude decrease in the X-ray luminosity of the nucleus
to be detected with respect to the faintest \ro\ measurement.  In fact, there
was no definite nuclear X-ray source to be seen in 2002, and the extreme
amplitude of the decline leads us to favor the tidal disruption interpretation
of the flare even if a low-luminosity AGN was present previously.
It is likely that a prior AGN accretion disk survived the tidal disruption event;
deeper exposures with \chandra\ should be able to detect its continuing
X-ray luminosity as an excess above the extrapolated $t^{-5/3}$ decay.

Occasional reobservation of these objects
is needed, if only to allow the tidal disruption 
hypothesis to be falsified by detecting renewed activity.
While some of the tidal debris should itself spread into a thin accretion disk
on the viscous time-scale, weak emission from this eventual AGN fuel
source is not expected to
dominate for several thousand years \citep{li02}.  In the likely event 
that all three X-ray sources studied here
continue to decline, deeper \chandra\ observations are 
required to track their luminosities and provide further observational
constraints on the decay curve that the theory of the fall-back phase predicts.

\acknowledgments

\clearpage
\begin{deluxetable}{lclrcccc}
\tabletypesize{\scriptsize}
\tablecaption{Summary of \chandra\ Observations \label{tbl1}}
\tablewidth{0pt}
\tablehead{
\colhead{Galaxy} &
\colhead{$z$} &
\colhead{Date} &
\colhead{$t_{\rm exp}$} &
\colhead{Photons} &
\colhead{$N_{\rm H}$(Gal)} &
\colhead{$f_{\rm X}(0.2-2.4$ keV)\tablenotemark{a}} &
\colhead{$L_{\rm X}(0.2-2.4$ keV)\tablenotemark{b}} \\
\colhead{} &
\colhead{} &
\colhead{} &
\colhead{(s)} & 
\colhead{$(0.2-5.0$ keV)} &
\colhead{(cm$^{-2}$)} &
\colhead{(ergs cm$^{-2}$ s$^{-1}$)} &
\colhead{(ergs s$^{-1}$)}
}
\startdata
RXJ1242.6--1119A point & 0.0510  & 2001 Mar  9 &  4477    & 18   & $3.7 \times 10^{20}$ 
                       & $(1.3 \pm 0.3) \times 10^{-14}$  & $(1.1 \pm 0.3) \times 10^{41}$ \\
NGC 5905 diffuse       & 0.0113  & 2002 Oct  4 &  9626    & 48   & $1.5 \times 10^{20}$
                       & $(1.5 \pm 0.3) \times 10^{-14}$  & $(4.4 \pm 0.9) \times 10^{39}$ \\
NGC 5905 point         & 0.0113  & 2002 Oct  4 &  9626    & $<8$ & $1.5 \times 10^{20}$
                       & $ <2.9 \times 10^{-15}$          & $< 9.1 \times 10^{38}$ \\
RXJ1624.9+7554 point   & 0.0639  & 2002 Sep 15 & 10086    &  4   & $3.9 \times 10^{20}$
                       & $(1.3 \pm 0.7) \times 10^{-15}$  & $(1.7 \pm 0.8) \times 10^{40}$ \\
\enddata
\tablenotetext{a}{Absorbed flux assuming a power-law of $\Gamma = 2.5 \pm 0.2$
for a point source, or a Raymond-Smith thermal plasma of log $T = 6.7\pm0.3$
and Solar abundances for a diffuse source.}
\tablenotetext{b}{Unabsorbed luminosity assuming $H_0 = 75$
km~s$^{-1}$~Mpc$^{-1}$.}
\end{deluxetable}

\begin{deluxetable}{lcccc}
\tabletypesize{\scriptsize}
\tablecaption{Optical Counterparts of \chandra\ Sources \label{tbl2}}
\tablewidth{0pt}
\tablehead{
\colhead{Source} &
\colhead{X-ray Position} &
\colhead{Photons} &
\colhead{Optical Position} &
\colhead{$R$} \\
\colhead{} &
\colhead{R.A. \hskip 0.4in  Decl.} &
\colhead{} &
\colhead{R.A. \hskip 0.4in  Decl.} &
\colhead{(mag)}
}
\startdata
RX J1242.6--1119A         & 12 42 38.56 --11 19 20.7  & 18 & 12 42 38.528 --11 19 20.44  & 14.1 \\
NGC 5905                  & 15 15 23.35  +55 31 02.2  & 48 & 15 15 23.320  +55 31 02.20  & ...  \\
RX J1624.9+7554           & 16 24 56.69  +75 54 55.6  &  4 & 16 24 56.648  +75 54 56.14  & 15.42 \\
CXOU J162434.1+755529     & 16 24 34.13  +75 55 29.7  & 32 & 16 24 34.154  +75 55 29.82  & 21.37 \\
CXOU J162442.7+755659     & 16 24 42.70  +75 56 59.1  & 12 & 16 24 42.793  +75 56 59.26  & 21.26 \\
CXOU J162501.6+755512     & 16 25 01.63  +75 55 12.5  & 12 & 16 25 01.705  +75 55 12.86  & 21.57 \\
\enddata
\tablecomments{Units of right ascension are hours, minutes, and seconds,
and units of declination are degrees, arcminutes, and arcseconds.}
\end{deluxetable}

\clearpage

\end{document}